\documentclass[prd,twocolumn,showpacs,superscriptaddress]{revtex4-1}

\usepackage[letterpaper,top=1.5cm,bottom=1.75cm,left=2cm,right=2cm]{geometry}
\usepackage{amssymb,amsmath,amsthm,mathptmx,latexsym,bm}
\usepackage{multirow,tabularx}
\usepackage[abs]{overpic}
\usepackage[colorlinks=true,linktocpage=true,linkcolor=blue,citecolor=blue]{hyperref}
\usepackage{fancyhdr}
\usepackage{float}
\newcommand{\be}{\begin{equation}}
\newcommand{\ee}{\end{equation}}
 \newcommand{\bea}{\begin{eqnarray}}
 \newcommand{\ena}{\end{eqnarray}}

\fancypagestyle{plain}

\newcolumntype{Y}{>{\centering\arraybackslash}X}

\begin{document}


\title{Thermodynamics and weak cosmic censorship conjecture with pressure in the rotating  BTZ black holes}

\author{Xiao-Xiong Zeng}
\affiliation{Department of Mechanics, Chongqing Jiaotong University, Chongqing 400074, China}
\affiliation{State Key Laboratory of Mountain Bridge and Tunnel Engineering, Chongqing Jiaotong University,\\ Chongqing 400074, China }
\author{Xin-Yun Hu}\email{xxzengphysics@163.com}
\affiliation{College of Economic and Management, Chongqing Jiaotong University,
 Chongqing 400074, China}

  \begin{abstract}
As  a charged spinning fermion drops into a charged rotating BTZ black hole, we investigate the laws of thermodynamics and  weak cosmic censorship conjecture with and without pressure respectively. For the case without pressure, the first law,  second law, as well as the weak cosmic censorship are found to be valid. While for the case with pressure, though the first law is still  valid,  the second law and the weak
cosmic censorship conjecture   are  found to be violable,  depending on the charge, angular momentum, AdS radius, and their variations.
In addition, in both cases, the  configurations  of  the extremal  black holes  are found to be stable  since the final states of the extremal   black holes  are still extremal black holes. While for the near-extremal black holes, their configurations are not stable.
  \end{abstract}
  \thispagestyle{fancy}
\maketitle
  \section{Introduction}

According to the singularity theorems proposed  by Penrose and Hawking \cite{Hawking:1969sw}, we know that the existence of a  singularity is  inevitable in Einstein's gravity.
   A singularity means  the invalidity of a classical gravity theory  so that it is not favorite by theoretical physicists.
In order to maintain the validity  of the classical gravity theory, Penrose  conjectured that  a singularity produced in
the gravitational collapse can not be observed by a distant observer
since it is hidden behind a horizon \cite{Penrose}, which is the so-called  weak cosmic censorship conjecture.  Now,  there exists no general  proof of this conjecture, so   we should check it one by one in each model of gravity. Around this topic, there are two strategies to check the weak cosmic censorship conjecture. On one hand,  one can check whether the gravitational collapse will end in a naked singularity without a horizon initially. In the
Einstein-Maxwell theory \cite{Crisford2017} and the Einstein-Maxwell-dilaton theory \cite{Goulart:2018ckh},  the weak cosmic censorship conjecture has been  found to be violated recently.
On the other hand, one also can check  whether the horizon of a charged/rotating black hole will  be destroyed after  a charged/rotating particle is absorbed. From the pioneering work of Wald \cite{Wald:1974ge}, there have been many investigations to check the weak cosmic censorship conjecture  with the test particle model \cite{Jacobson:2009kt,Gao:2012ca,Barausse:2010ka,Colleoni:2015afa,Wang:2019jzz,Eperon:2019viw,Gim:2018axz,Liang:2018wzd,Isoyama:2011ea}. Among them, the  backreaction and self force effect  were found to be important to the weak cosmic censorship conjecture \cite{Barausse:2010ka}. It has been found that particles that
could overspin the black hole \cite{Jacobson:2009kt} would not be able to fall into the horizon as these effects are taken into account.

The test particle model is prevail for the following two advantages. Firstly, besides the  weak cosmic censorship conjecture, it also can be used to investigate the first and second laws of thermodynamics \cite{Gwak:2015sua}. Secondly,  it also can be used to investigate the laws of thermodynamics and   weak cosmic censorship conjecture in the extended phase space with pressure \cite{Gwak:2017kkt}.
Recently, there have been some investigations on the laws of thermodynamics and   weak cosmic censorship conjecture in the extended phase space with pressure \cite{Zeng:2019jta,Wang:2019dzl,Han:2019kjr,Zeng:2019jrh,Han:2019lfs,Zeng:2019aao,Chen:2019nsr,Zeng:2019hux,He:2019fti,Mu:2019bim,Chen:2019pdj}. It has been found that the second law will be violated as the pressure is considered though the first law and the  weak cosmic censorship conjecture are valid. In addition, the configuration of the black hole was found to be not changed after a particle is absorbed in the extended phase space with pressure.  The pressure of the  black hole thus plays a crucial role when we discuss the thermodynamic laws and   weak cosmic censorship conjecture.

In this paper, we will employ the test particle model to investigate the thermodynamic laws and weak cosmic censorship conjecture of a charged rotating BTZ black hole with or without pressure. Our motivation is twofold. On one hand, we want to explore whether the topology of the spacetime  affect the thermodynamic laws and weak cosmic censorship conjecture. Previously, most investigations were limited to the spherically symmetric black holes \cite{Zeng:2019jta,Wang:2019dzl,Han:2019kjr,Zeng:2019jrh,Han:2019lfs,Zeng:2019aao,Chen:2019nsr,Zeng:2019hux,He:2019fti,Mu:2019bim,Chen:2019pdj}. However for the rotating black hole, the topology of the spacetime is cylindrically symmetric for the existence of a angular momentum. We want to explore how the angular momentum  affect the thermodynamic laws and weak cosmic censorship conjecture. On the other hand, we want to explore whether the high order corrections to the mass of the particle affect the validity of the  weak cosmic censorship conjecture.
Recently, by  applying the  Wald formalism, the weak cosmic censorship
conjecture  was  found to be  valid for  the non-extremal black holes  \cite{Wald2018,Sorce:2017dst,Ge:2017vun,An:2017phb} as the second order variation of the
mass of the black hole was considered. The result is different from the previous investigations where   the second order correction was not taken into account \cite{Jacobson:2009kt}. With the test particle model in this paper, we want to explore whether the  weak cosmic censorship
conjecture are valid as the second order correction to the mass is considered with or without pressure. As a result, we find for the case without pressure, the  weak cosmic censorship
conjecture is valid, while for the case with pressure, the weak
cosmic censorship conjecture  is violable,  depending on the charge, angular momentum, AdS radius, and their variations.

This paper is outlined as follows. In section \ref{2},  we introduce the thermodynamic quantities of the charged rotating BTZ black hole. In section \ref{3}, we obtain the energy-momentum relation of a charged spinning particle as  it drops into the  charged rotating BTZ black hole. In section \ref{4} and section \ref{5},
 the first law, second law as well as  weak cosmic censorship conjecture are investigated  with and without pressure respectively. Especially, we investigate how the high order corrections to the  mass   of the particle affect the weak cosmic censorship conjecture.
 Section \ref{6} is devoted to our conclusions.

  \section{Reviews of the   charged rotating BTZ black holes }
  \label{2}
The charged rotating BTZ black hole is produced by the
 action \cite{Banados:1992wn}
\be
I=\frac{1}{16\pi G}\int d^{3}x\sqrt{-g}\left( R-2\Lambda -4\pi GF_{\mu \nu
}F^{\mu \nu }\right) ,  \label{action}
\ee
with
\be
F_{\mu \nu }=A_{\nu ,\mu }-A_{\mu ,\nu },
\ee
where  $G$ is the gravitational constant in three dimension, $\Lambda$ is the
cosmological constant that relates  to the AdS radius as $\Lambda=-1/l^2$, $R$ is the Ricci scalar, $g$ is determinant of the
metric tensor, and $A_{\mu }$ is the electrical potential.
The  line element in Schwarzschild coordinate is given by
\be
ds^{2}=-F\left( r\right) dt^{2}+F^{-1}\left( r\right) dr^{2}+r^{2}\left(
d\phi -N^{\phi}dt\right) ^{2},  \label{BTZ}
\ee
where%
\be
F(r)=-M+\frac{r^{2}}{l^{2}}-\frac{1}{2}Q^{2}\ln \left( \frac{r}{l}\right) +%
\frac{J^{2}}{4r^{2}},  \label{frQ}
\ee
\be
N^{\phi}=\frac{J}{2r^{2}},
\ee
in which $M$ is the mass, $Q$ is the electric charge, and $J$ is the angular momentum of the
black hole. The nonvanishing component of the vector potential is
\be
A_{t}=Q\ln (\frac{r}{l}).
\ee
With the equation $F(r)=0$, we can obtain two solutions and the largest solution is the   event horizon, labeled as $r_+$.   The mass of the black hole is%
\be  \label{mr}
M (Q,J,r_+)=\frac{r_{+}^{2}}{l^{2}}-\frac{1}{2}Q^{2}\ln \left( \frac{r_{+}}{l%
}\right) +\frac{J^{2}}{4r_{+}^{2}}.
\ee
According to the area of the horizon, the entropy can be expressed as
\be  \label{sr}
S=4\pi r^{+},
\ee
where we have used the unit $8G=1$.
The Hawking
temperature of this black hole is
\be
T=F^{\prime }(r_{+})/4\pi=\frac{1}{4\pi }\left( \frac{2r_{+}}{l^{2}}-\frac{Q^{2}}{2r_{+}}-\frac{%
J^{2}}{2r_{+}^{3}}\right).  \label{tr}
\ee
At the horizon, the angular velocity and the electric potential are
\be  \label{or}
\Omega_+ =\frac{J}{2r_{+}^{2}},
\ee
\be  \label{phi}
\Phi=-Q\ln \left( \frac{r_{+}}{l}\right).
\ee
From Eqs.(\ref{mr}), (\ref{sr}), (\ref{tr}), (\ref{or}), and (\ref{phi}), we can get the first law of thermodynamics
\be \label{thefirstlaw}
dM=T dS+ \Omega_+ dJ+\Phi dQ.
\ee
In Eq.(\ref{thefirstlaw}), the cosmological parameter is regarded as a constant, the phase space of the thermodynamics is the normal phase space. Recent investigations have  shown that the cosmological parameter can be treated as a dynamical variable.
 In this case,
 the first law of thermodynamics  changes into
\be \label{firstlaw1}
dM=T dS+ \Omega_+ dJ+\Phi dQ+VdP,
\ee
in which $P$ is the pressure and $V$ is the volume of the thermodynamic system, which are defined as respectively \cite{Sadeghi:2015out,Zou:2014gla}
\be \label{pp}
P=\frac{1}{8 \pi  l^2},~~
\ee
\be
V=8 \pi  r_+^2-2 \pi  l^2 Q^2. \label{vv}
\ee
In Eq.(\ref{firstlaw1}), the mass $M$ is interpreted not as the internal energy but the enthalpy. The thermodynamic phase space is the extended phase space in this case.

\section{Energy-momentum relation of a charged spinning fermion }\label{3}

 In this section, we are going to find the energy-momentum relation of a charged spinning fermion as it drops into a  charged rotating BTZ black hole.  We will employ the following   Dirac equation for electromagnetic field \cite{Ejaz:2013fla,Zeng:2008zzc}
\be
i\gamma ^{\mu }\left( \partial _{\mu }+\Omega _{\mu }-\frac{i}{\hbar }
eA_{\mu }\right) \Psi -\frac{\mu }{\hbar }\Psi =0, \label{diraceq}
\ee
where
\bea
\Omega _{\mu } &=&\frac{i}{2}\Gamma _{\mu }^{\alpha \beta }\Sigma _{\alpha
\beta }, \\
\Sigma _{\alpha \beta } &=&\frac{i}{4}\left[ \gamma ^{\alpha },\gamma
^{\beta }\right] ,\\
\Omega _{\mu }&=&\frac{-1}{8}\Gamma _{\mu }^{\alpha
\beta }\left[ \gamma ^{\alpha },\gamma ^{\beta }\right].
\ena
For a rotating BTZ black hole, its event horizon does not coincide with
the infinite red-shift surface due to the existence
of rotation. There is an ergosphere between the event horizon and infinite red-shift surface, and matter  near the horizon will be dragged inevitably by the gravitational
field. To investigate the motion of the spinning fermion conveniently, we will make the following dragging coordinate transformation
\be
\psi=\phi- {\Omega} t,
\ee
in which we have defined the dragging angular velocity
\be
\Omega=N^{\phi}=\frac{J}{2r^{2}}.
\ee
In this case, the metric in  Eq.(\ref{BTZ}) becomes as
\be
ds^{2}=-F\left( r\right) dt^{2}+F^{-1}\left( r\right) dr^{2}+r^{2}d\psi ^{2}.
\label{btz1}
\ee
We set the  $\gamma ^{\mu }$ matrices as
\bea
\gamma ^{\mu}=\left( -iF^{-\frac{1}{2}}\sigma ^{2},F^{\frac{1}{2}}\sigma ^{1},\frac{1}{r}\sigma ^{3}\right) ,
\ena
where we have employed the Pauli sigma matrices
\be
\sigma ^{1}=\left(
\begin{array}{cc}
0 & 1 \\
1 & 0
\end{array}
\right),~\sigma ^{2}=\left(
\begin{array}{cc}
0 & -i \\
i & 0
\end{array}
\right) ,~\sigma ^{3}=\left(
\begin{array}{cc}
1 & 0 \\
0 & -1
\end{array}
\right).
\ee
For a fermion with spin 1/2, the wave function has two states, namely spin up  and spin down. In this paper, we  focus on the
spin up state for the case of spin down  is similar. We  will adopt the following ansatz for the solution
\begin{eqnarray}
\Psi _{\uparrow } &=&\left(
\begin{array}{c}
C\left( t,r,\psi \right)  \\
D\left( t,r,\psi \right)
\end{array}%
\right) e^{\frac{i}{\hbar }I_{\uparrow }\left( t,r,\psi \right) }.
\label{spinup}
\end{eqnarray}%
Inserting Eq.(\ref{spinup})  into   Eq.(\ref{diraceq}),
we have the following two simplified equations
\bea
D%
\left[ \sqrt{F}\partial _{r}I_{\uparrow }-\left( \frac{1}{\sqrt{F%
}}\partial _{t}I_{\uparrow }-\frac{1}{\sqrt{F}}e A_t \right) \right]\nonumber\\
+C\left( \mu +\frac{1}{r}\partial _{\psi }I_{\uparrow } \right) =0,
\ena
\bea
C\left[ \sqrt{F}\partial _{r}I_{\uparrow }+\left( \frac{1}{\sqrt{%
F}}\partial _{t}I_{\uparrow } -\frac{1}{\sqrt{F}}e A_t \right) \right]\nonumber \\
+D\left( \mu -\frac{1}{r}\partial _{\psi
}I_{\uparrow }\right) =0,
\ena
in which we only take the leading order of $\hbar$. To assure $C$ and $D$ have
non-trivial solutions, the determinant of the
coefficient matrix should vanish, leading to
\bea\label{meq1}
\frac{1}{r^{2}}\left( \partial _{\psi }I_{\uparrow }
\right) ^{2}-\mu ^{2}+\left( \sqrt{F}\partial _{r}I_{\uparrow }\right) ^{2}
\nonumber \\-\left( \frac{1}{\sqrt{F}}\partial _{t}I_{\uparrow } -\frac{1}{\sqrt{F}}e A_t \right) ^{2}=0.
\ena
In the dragging coordinate system, the action $I_{\uparrow }$ can be expressed as
\begin{equation} \label{action1}
I_{\uparrow }=-(\omega-j \Omega_+ )t +L\psi +W\left( r\right),
\end{equation}%
where $\omega $, $L$, and $\Omega_+$ are fermion's energy, angular momentum, and angular velocity at the horizon.
 Putting Eq.(\ref{action1}) into Eq.(\ref{meq1}), we obtain
\begin{equation}
\partial _{r}W\left( r\right) =\pm \frac{1}{F}\sqrt{\left( \omega -j \Omega_+ +eA_t \right) ^{2}+F\left( \mu ^{2}-\frac{L^{2}}{r^{2}}%
\right) }.
\end{equation}%
 We want to investigate the thermodynamics and weak  cosmic  censorship  conjecture, so we are interested only in the location at the horizon. In this case,
 the radial momentum of the particle can be expressed as
\bea
|p^r_+|=g^{rr}\partial _{r}W\left( r\right)\mid_{r_+}
=\omega+eA_t(r_+)-j \Omega_+, \label{emrq}
\ena
in which $p^r_+$ is the radial momentum at the horizon. As stressed in \cite{ref8},  in order to assure the particle is absorbed  in the positive flow of time,  a positive sign should be endowed in front of  $|p^r_+|$.


\section{Thermodynamics and  weak cosmic  censorship  conjecture without pressure }\label{4}
\label{3}
In this section, we will employ  Eq.(\ref{emrq})  to investigate the thermodynamics and cosmic  censorship  conjecture without pressure, namely the cosmological parameter is a constant.
As a
 charged spinning fermion is absorbed by the charged rotating BTZ black hole, the variation of the internal energy, angular momentum  and charge  of the black hole satisfy
\be
\omega=dM, j=dJ, e=dQ,
\ee
where we have  used  the energy conservation, angular momentum conservation  and charge conservation. In this case,  the energy-momentum relation in Eq.(\ref{emrq}) can be rewritten as
\be \label{emrq1}
dM=\Phi dQ+p^r_++j \Omega_+.
\ee
As the charged spinning  fermion is absorbed,  the event horizon of the black hole  will change from $r_{+}$ to $r_{+}+dr_{+}$. The new event
horizon satisfies also the equation of horizon,  $F(r_++dr_+)=0$. In other words,   there is always a relation
\bea \label{dff}
dF&=&F(r_++dr_+)-F(r_+) \nonumber \\
&=&\frac{\partial F}{\partial M}dM+\frac{\partial F}{\partial Q}dQ+\frac{\partial F}{\partial r_+}dr_+ +\frac{\partial F}{\partial J}dJ=0.~~~~
\ena
Inserting Eq.(\ref{emrq1}) into Eq.(\ref{dff}), we find $dM$,  $dQ$ and $dJ$ are deleted. So  we can solve $dr_+$  directly, which is
 \be\label{dr}
dr_+=-\frac{2 l^2 p^r_+ r_+^3}{J^2 l^2+l^2 Q^2 r_+^2-4 r_+^4}.
\ee
Making use of Eq.(\ref{sr}),  the variation of entropy  can be expressed as
\be \label{dsr}
dS=-\frac{8\pi l^2 p^r_+ r_+^3}{J^2 l^2+l^2 Q^2 r_+^2-4 r_+^4}.
\ee
Combining Eqs.(\ref{tr}) and (\ref{dsr}), we find there is a relation
\be
T dS=p^r_+.\label{tspr}
\ee
In this case, the internal energy in Eq.(\ref{emrq1}) can be reexpressed as
\begin{equation} \label{eq:dm1}
dM=TdS+\Phi dQ+ \Omega_+ dJ,
\end{equation}
which is nothing but the first law of   black hole thermodynamics. In other words, as a charged  spinning fermion  is absorbed by the charged rotating BTZ black hole, the first law
is valid without pressure.

We also can  check the validity of the second law of  thermodynamics, which states that the entropy of a black hole never decrease in the clockwise direction. As a charged spinning fermion drops into  the black hole, the entropy of the black hole should increase according to the second law of the thermodynamics. We will adopt Eq.(\ref{dsr}) to check whether this is true for the charged rotating BTZ black hole.

For an extremal charged rotating BTZ  black hole, we know that the temperature of the black hole vanishes at the horizon, and  the inner horizon and outer horizon are coincident.
  With Eq.(\ref{tr}), we can get the radius of the extremal charged rotating BTZ  black hole
\be\label{re}
r_e=\frac{\sqrt{\sqrt{l^2 \left(16 J^2+l^2 Q^4\right)}+l^2 Q^2}}{2 \sqrt{2}}.
\ee
Substituting it  into Eq.(\ref{dsr}), we find the variation of the entropy is   divergent,  which  implies that the second law for the extremal BTZ black hole is meaningless  since the thermodynamic system is  a zero temperature system.

\begin{figure}[H]
\centering
\includegraphics[scale=0.55]{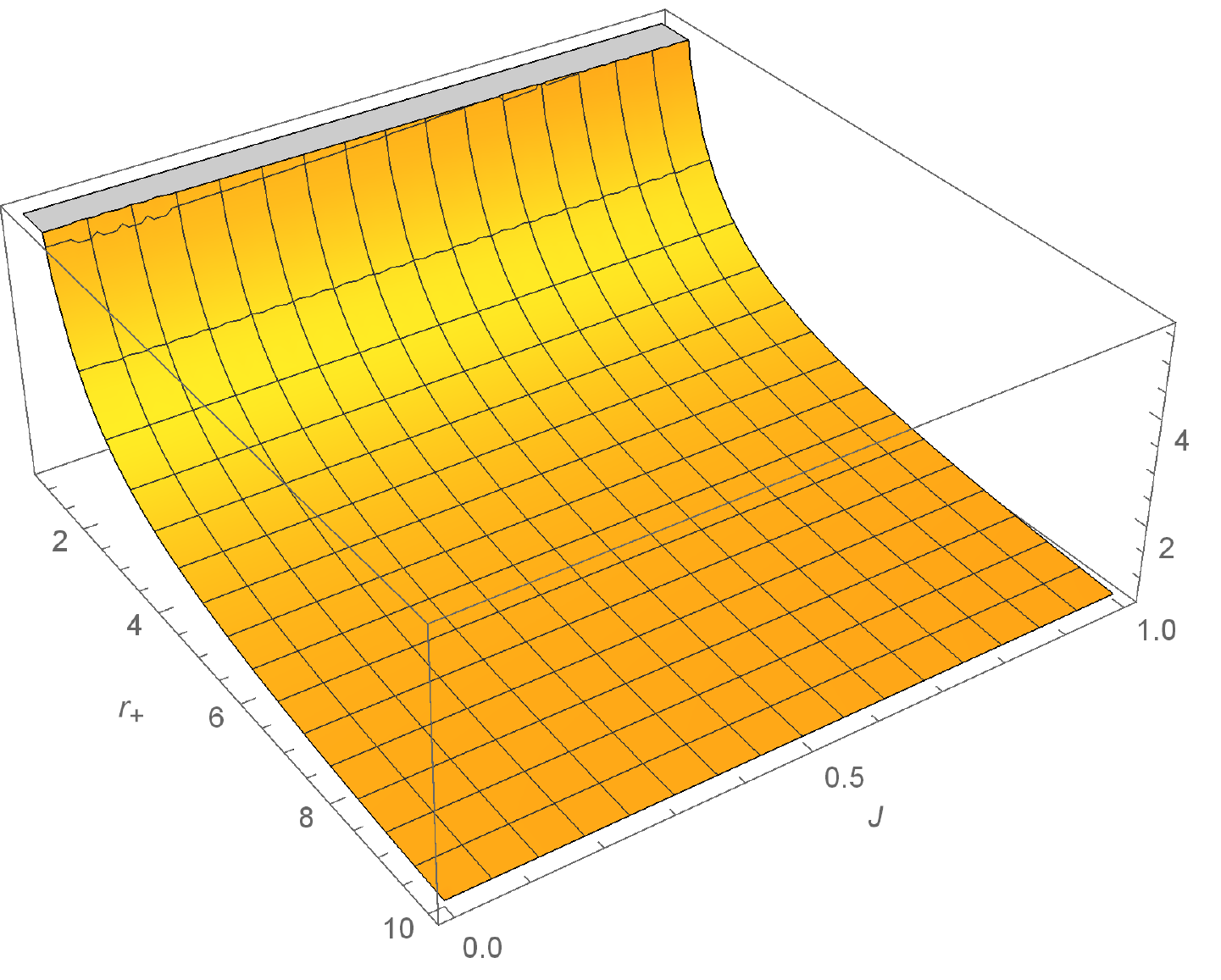}
 \caption{\small The values  of $dS$ for different  $J$ and $r_+$  for the case $p^r_h = l = Q=1$.   } \label{fig1}
\end{figure}

For a non-extremal charged rotating  BTZ black hole, the  temperature is larger then zero, and the variation of the entropy is nonvanishing. For different  $J$ and $r_+$, the variation of entropy is plotted
in Fig. \ref{fig1}. From this figure, we can see that
  $dS$   is positive always. Therefore, the second law of thermodynamics is valid in the normal phase space without pressure.

It should be stressed that
as the values of  $p^r_h, l, Q$ change, the configuration of Fig. \ref{fig1} will vary correspondingly. But no matter how they change, the conclusion that  $dS$   is positive will not change.

Next, we turn to   the weak
cosmic censorship conjecture, which states that the observer located at future null infinity  can not observe the singularity of a spacetime  for the singularity is covered by the event horizon.   So the validity of the weak
cosmic censorship conjecture means the existence of
an event horizon. As a charged spinning fermion is absorbed by a charged rotating BTZ  black hole, we intend to check whether there is an event horizon. Namely, whether the equation  $F(r)=0$ has solutions at the final state.
We will judge it by investigating the
 minimum value  of  $F(r)$, located at the radial coordinate $r_{m}$. When $F(r_m)>0$, there is not a horizon while when $F(r_m)\leq0$, there are horizons always.   At $r_{m}$, we have
\bea
&F(r)|_{r=r_{m}}\equiv F_{m}=\delta \leq 0,\label{cn1} \\
&\partial_{r}F(r)|_{r=r_{m}}\equiv F'_{m}=0.\label{cn2}
\ena
 For an extremal rotating BTZ black hole, $\delta=0$,  $r_+$ and   $r_m$ are coincident. For a  near-extremal BTZ black hole, $\delta$ is a small quantity, $r_{m}$ is distributed between the outer horizon and inner horizon.

In the normal phase space, the initial state of the charged rotating BTZ  black hole is labelled by $M, Q, J$.   As a charged spinning fermion drops into the BTZ black hole, the mass, charge and angular momentum of the black hole will change into $M+dM, Q+dQ,  J+dJ$ respectively. Correspondingly, the radial coordinate $ r_{m}$ and event horizon $r_{+}$ will change into $ r_{m}+dr_{m}$, $r_{+}+dr_{+}$. There is also a shift for the function $F(r)$, which is defined by
 \bea \label{ddfm}
dF_{m}&=&F(r_{m}+dr_{m})-F(r_{m})\nonumber \\
&=&\left(\frac{\partial F_{m}}{\partial M}dM+\frac{\partial F_{m}}{\partial Q}dQ+\frac{\partial F_{m}}{\partial J}dJ\right)\nonumber \\
&=&\frac{{dJ} J}{2 r_m^2}-{dM}-{dQ} Q \log \left(\frac{r_m}{l}\right),
\ena
where we have used the condition $F'_{m}=0$ in Eq.(\ref{cn2}).
Our next step is to find the last result of Eq.(\ref{ddfm}) in different cases.

For an extremal black hole,   $r_+$ and   $r_m$ are coincident,   Eq.(\ref{emrq1}) thus is applicable. Inserting Eq.(\ref{emrq1}) into Eq.(\ref{ddfm}), we find $dQ, dM, dJ$ are deleted. In this case, Eq.(\ref{ddfm}) can be simplified lastly as
\be
 dF_{m}=-p^r_+. \label{dfml}
\ee
In \cite{Gwak:2017kkt}, it has been claimed that $ dF_{m}$ is negative on the assumption that $p^r_+$ is positive. However,  the temperature of the extremal black hole vanishes, so $p^r_+$  vanishes too according to Eq.(\ref{tspr}).
Therefore $dF_{m}=-p^r_+=0$. In other words, the extremal rotating BTZ  black hole will not change its configuration for its final state is still an extremal rotating BTZ black hole.

For a near-extremal charged rotating BTZ  black hole, Eq.(\ref{emrq1})  is not applicable  at $r_m$ for it holds only at the event  horizon. With the condition $r_{+}=r_m+\epsilon$, we can expand Eq.(\ref{emrq1}) at
$r_m$, which leads to
\bea
dM&=&\frac{ {dJ} J}{2 r_m^2}- {dQ} Q \log \left(\frac{r_m}{l}\right)-\left(\frac{ J^2}{2 r_m^3}-\frac{2   r_m}{l^2}+\frac{  Q^2}{2 r_m}\right){dr_m}\nonumber \\
&+& \left(-\frac{ {dJ} J}{r_m^3}-\frac{ {dQ} Q}{r_m}+\frac{3  {dr_m} J^2}{2 r_m^4}+\frac{2  {dr_m}}{l^2}+\frac{ {dr_m} Q^2}{2 r_m^2}\right)\epsilon\nonumber\\
&+&O\left(\epsilon^2\right), \label{exm}
\ena
where we have used Eq.(\ref{tspr}). Substituting Eq.(\ref{exm}) into Eq.(\ref{ddfm}), we get lastly
\bea
dF_{m}&=&\frac{ {dr_m} \left(J^2-{4 r^4_m}/{l^2}+Q^2 r_m^2\right)}{2 r_m^3}\nonumber\\
&+& \left(\frac{ {dJ} J+ {dQ} Q r_m^2}{r_m^3}+ {dr_m} \left(-\frac{3 J^2+Q^2 r_m^2}{2 r_m^4}-\frac{2}{l^2}\right)\right)\epsilon\nonumber\\
&+&O\left(\epsilon^2\right). \label{dfn}
\ena
To simplify  Eq.(\ref{dfn}), we are going to find $dJ$  and $J$. According to  Eq.(\ref{cn2}),  we find
 \be
 J=\frac{\sqrt{4 r_m^4-l^2 Q^2 r_m^2}}{l}. \label{jn}
 \ee
 In addition,  at  $r_m+dr_m$, there is also a relation
\be
\partial_{r} F(r)|_{r=r_m+dr_m}
=F'_{m}+dF'_{m}=0, \label{pf}
\ee
which implies
\be
dF'_{m}=\frac{\partial F'_{m}}{\partial Q}dQ+\frac{\partial F'_{m}}{\partial J}dJ+\frac{\partial F'_{m}}{\partial r_{m}}dr_{m}=0.
\ee
Solving this equation, we   get
\be
dJ=\frac{r_m \left(-{dQ} l^2 Q r_m-{dr_m} l^2 Q^2+8 {dr_m} r_m^2\right)}{l \sqrt{4 r_m^4-l^2 Q^2 r_m^2}}.  \label{djn}
\ee
 Substituting  Eqs.(\ref{jn}), (\ref{djn}) into Eq.(\ref{dfn}), we obtain lastly
 \bea
dF_{m}=O\left(\epsilon^2\right).
\ena
In \cite{Gwak:2017kkt}, it was claimed that $O\left(\epsilon^2\right)$ is the high order terms of $\epsilon$, so it can be ignored. In fact, $\delta$ is also a small quantity, we can not ignore the contribution of the $O\left(\epsilon^2\right)$ term. Next, we will find the coefficient of $O\left(\epsilon^2\right)$  and the value of  $\epsilon$ to discuss which is smaller.

\begin{figure}[H]
\centering
\includegraphics[scale=0.55]{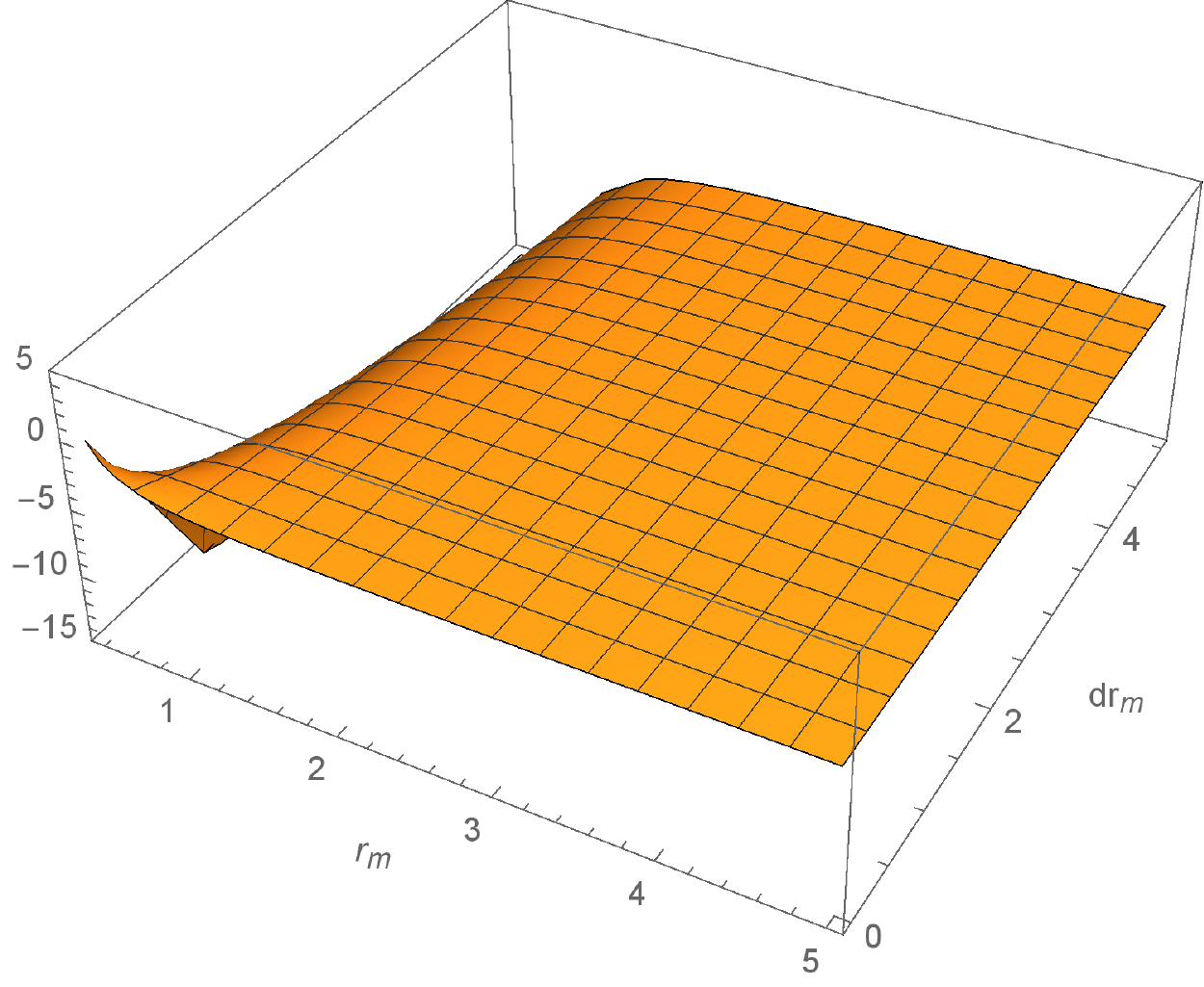}
 \caption{\small The values  of $\bar F $ for different  $r_m$ and $dr_m$  for the case $Q=l=1$, $dQ=0.5$.   } \label{fig3}
\end{figure}

To obtain the value of  $\delta$,  we expand  $F(r_+)$ at $r_m$ to the second order, leading to
\bea
0=F(r_+)&=&\frac{2 r_m^2}{l^2}-\frac{1}{2} Q^2 \log \left(\frac{r_m}{l}\right)-M-\frac{Q^2}{4}\nonumber\\
&+& \left(\frac{4}{l^2}-\frac{Q^2}{2 r_m^2}\right)\epsilon^2+O\left(\epsilon^3\right). \label{eef}
\ena
That is to say,
\be \label{deltaf}
\delta=F(r_{m})=-\left(\frac{4}{l^2}-\frac{Q^2}{2 r_m^2}\right)\epsilon^2-O\left(\epsilon^3\right).
\ee
In addition, to the second order, $dF_{m}$ can be expressed as
\be
dF_{m}=\frac{Q   \left( {dQ} r_m- {dr_m} Q\right)}{r_m^3}\epsilon^2+O\left(\epsilon^3\right).
\ee
In this case, we have
\bea
F(r_m+dr_m)&=&F(r_m)+dF_m\nonumber\\
&=& \left(\frac{Q \left( {dQ} r_m- {dr_m} Q\right)}{r_m^3}-\frac{4}{l^2}+\frac{Q^2}{2 r_m^2}\right)\epsilon^2.
\ena
We define $F(r_m+dr_m)/\epsilon^2$ as $\bar F$. If  the value of $\bar F$ is positive, the weak
cosmic censorship conjecture is violated, and if it is negative, the weak
cosmic censorship conjecture is valid. Here, we will obtain $\bar F$ numerically. For the case $Q=l=1$, $dQ=0.5$,  the value of $\bar F$  is shown in  Fig. \ref{fig3}. We find that  $\bar F$ is negative, which is independent on the value of $r_m, dr_m$.  In fact, as $ dQ, dr_m$ is smaller than $Q, r_m$, we find $\bar F$  is negative always as we vary the values of $Q, dQ, l, dr_m$,
 indicating  that the weak
cosmic censorship conjecture is  valid for the near-extremal black holes in the normal phase space without pressure. Our result is consistent with that in \cite{Wald2018,Sorce:2017dst}, where the high order corrections to the mass were considered. 

In addition, based on Eq.(\ref{re}), we can obtain the radius of the extremal black hole as the values of  $J, Q, l$ are given. The inial value of $r_m$ in   Fig. \ref{fig3} thus should be a little smaller than
the radius of the extremal black hole. As  $r_m$ approaches to the radius of the extremal black hole, we find $\bar F$ approaches to zero, which is consistent with the result in Eq.(\ref{dfml}).

\section{Thermodynamics and weak  cosmic  censorship  conjecture with pressure}\label{5}

In the extended phase space with pressure, the cosmological parameter is not a constant but a state parameter of thermodynamic system, interpreted as the pressure $P$.   In addition, the mass $M$ is not the internal energy but the  enthalpy, which relates to the internal energy   as
\be \label{mupv}
M=U+PV.
\ee
As a charged spinning fermion drops into the charged rotating BTZ black hole, the energy, charge, and angular momentum  are supposed to be conserved. Namely the energy, charge, and angular momentum  of the fermion equal to the varied energy, charge, and angular momentum  of the black hole, which implies
\be
\omega=dU=d(M-PV), e=dQ, j=dJ.
\ee
 The energy in Eq.(\ref{emrq}) changes correspondingly into
\be \label{due}
dU=\Phi dQ+p^r_++j \Omega_+.
\ee
The absorbed fermion will change  the location of the event horizon of the black hole. Correspondingly, the function $F(r)$ will be changed too, that is
\be  \label{dfei}
dF=\frac{\partial F}{\partial M}dM+\frac{\partial F}{\partial Q}dQ+\frac{\partial F}{\partial r_+}dr_+ +\frac{\partial F}{\partial J}dJ+\frac{\partial F}{\partial l}dl=0.~~~~
\ee
Different from the case without pressure, $l$ is now a variable in the extended phase space with pressure.

 In addition, with Eq.(\ref{mupv}), Eq.(\ref{due}) can be expressed as
\be \label{dme}
dM-d(PV)=\Phi dQ+p^r_++j \Omega_+.
\ee
Substituting  $dM$ in Eq.(\ref{dme}) into Eq.(\ref{dfei}), we can delete  $dJ$, $dQ$, $dl$, and $dM$. In this case, there is only a  relation between $p^r_+$  and  $dr_{+}$,  which is
 \be \label{dre}
dr_ {+}=-\frac{r_+^3 \left( {dV}+8 \pi  l^2 p^r_+\right)}{4 \pi  \left(J^2 l^2+l^2 Q^2 r_+^2-4 r_+^4\right)}.
\ee
So, the variation of entropy   of the black hole can be  written  as
\be\label{dse}
dS=-\frac{r_+^3 \left( {dV}+8 \pi  l^2 p^r_+\right)}{J^2 l^2+l^2 Q^2 r_+^2-4 r_+^4}.
\ee
Combining  Eqs.(\ref{tr}),  (\ref{pp}), and  (\ref{dse}), we find
\be \label{pr}
T dS-PdV=p^r_+.
\ee
The internal energy in Eq.(\ref{due}) thus can be rewritten as
\be  \label{du}
dU=\Phi dQ +j \Omega_+ + T dS-PdV.
\ee
 Moreover, from Eq.(\ref{mupv}), we can get
\be  \label{dm1}
dM=d U+P dV+VdP.
\ee
Substituting Eq.(\ref{dm1}) into Eq.(\ref{du}), we find
\be
dM=TdS+\Phi dQ+VdP+ \Omega_+ dJ,
\ee
which is consistent with that in Eq.(\ref{firstlaw1}). That is, as a charged spinning fermion is absorbed by the charged rotating BTZ black hole, the first law of thermodynamics holds in the extended phase space with pressure.

\begin{figure}[H]
\centering
\includegraphics[scale=0.55]{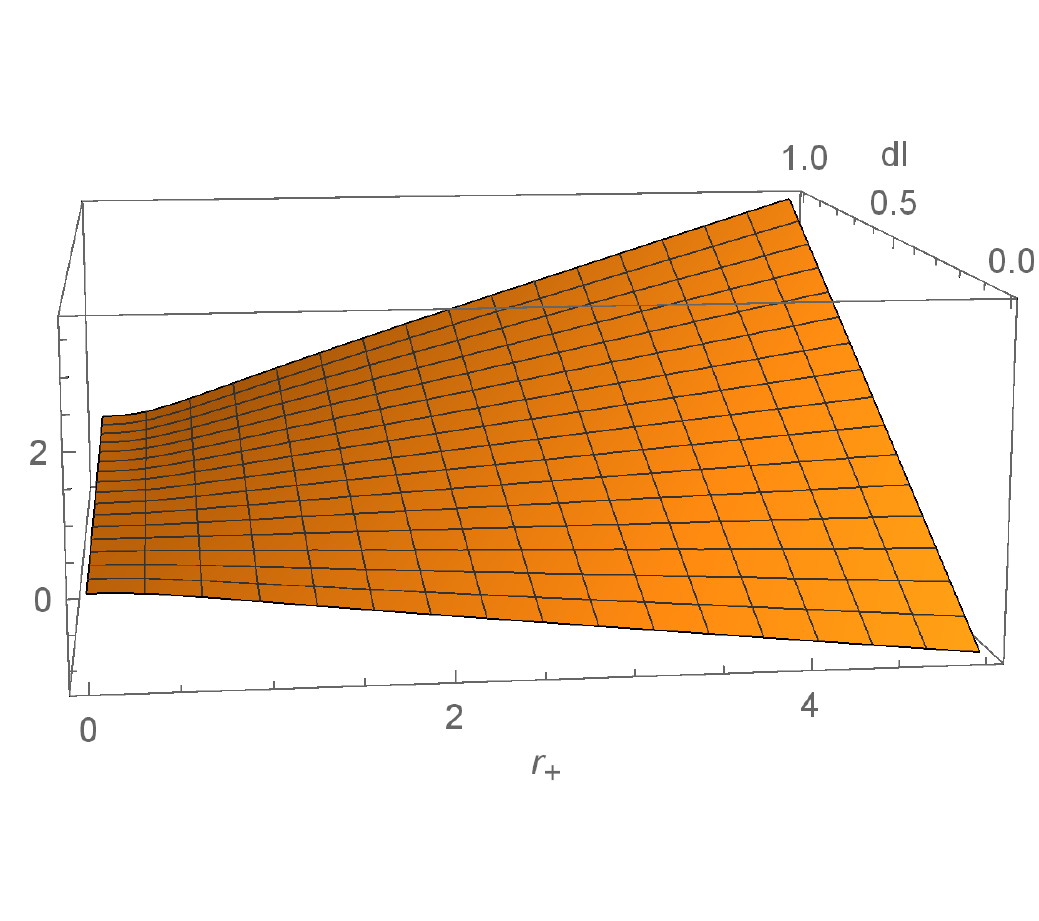}
 \caption{\small The  relations among $dS$  and $r_+, dl$ for   the case $J = l = 1$, $Q=2, dQ=0.5$. } \label{fig2}
\end{figure}

Based on Eq.(\ref{dse}), we also can discuss the second law of thermodynamics with pressure. As $dV$  is replaced with the help of Eq.(\ref{pp}), Eq.(\ref{dse}) changes into
\be \label{dsse}
dS=\frac{{dl} Q^2 r_+^3+{dQ} l Q r_+^3-2 l r_+^3}{l \left(J^2+Q^2 r_+^2\right)}.
\ee
We can see clearly that $dS$ may be negative or positive for the existence of terms  $dl, dQ$, which is different from that without pressure. For the case $J = l = 1$, $Q=2, dQ=0.5$, we give the relations among $dS$ and  $r_+, dl$, which is shown in Fig.  \ref{fig2}. It is obvious that $dS$ is negative for some values of $r_+, dl$. As  the values of  $J, l, Q, dQ$  change,  the configuration of Fig. \ref{fig2} will change too. However, the conclusion that $dS$ may be negative will not change. In other words, the second law in the extended phase space with pressure may be violated, depending on the values of $l, Q$ and their variations.

Next, we focus on investigating the   weak
cosmic censorship conjecture with pressure  on the basis of   Eqs.(\ref{cn1}) and (\ref{cn2}).
As a charged spinning fermion is absorbed by the charged rotating BTZ black hole,  the mass $M$, charge $Q$,  angular momentum $J$, and AdS radius $l$   will change into $M+dM, Q+dQ, J+dJ, l+dl$. Correspondingly, the  radial coordinate and event horizon will change into $ r_{m}+dr_{m}$, $ r_{+}+dr_{+}$.
In this case, the shift of $F(r_m)$ can be written as
\be  \label{eedf}
dF_m=\left(\frac{\partial F_{m}}{\partial M}dM+\frac{\partial F_{m}}{\partial Q}dQ+\frac{\partial F_{m}}{\partial l}dl+\frac{\partial F_{m}}{\partial J}dJ\right),
\ee
where we have used  Eq.(\ref{cn2}).
Next, we focus on  finding the last result of Eq.(\ref{eedf}).

For the extremal rotating  BTZ black hole, the horizon is located at $r_m$.
The energy of the particle in Eq.(\ref{due}) is valid. Substituting Eq.(\ref{due}) into  Eq.(\ref{eedf}), we find
\be
dF_m
=\frac{ {dr_{m}} \left(J^2-{4 r_m^4}/{l^2}+Q^2 r_m^2\right)}{2 r_m^3}. \label{eedf2}
\ee
Substituting Eq.(\ref{jn}) into  Eq.(\ref{eedf2}), we find
\be \label{dfee}
dF_m=0.
\ee
 That is, as a charged spinning fermion drop into an extremal charged rotating BTZ  black hole, the black hole  stays at its initial state so that  its configuration will not be changed, which is the same as that without pressure.

For the near-extremal rotating BTZ black hole,  the energy of the particle in Eq.(\ref{due}) is not valid for it is applicable only at the event horizon. But we can expand it near the lowest point with $r_+=r_m+\epsilon$. To the first order, we get
\bea
dM&=&\left(\frac{{dJ} J}{2 r_m^2}-\frac{2{dl} r_m^2}{l^3}+\frac{{dl} Q^2}{2 l}-{dQ} Q \log \left(\frac{r_m}{l}\right)\right)\nonumber\\
&+&\left(\frac{2 {dr_m} r_m}{l^2}-\frac{{dr_m} Q^2}{2 r_m}-\frac{{dr_m} J^2}{2 r_m^3}\right)\nonumber\\
&+& \left(-\frac{{dJ} J}{r_m^3}-\frac{4 {dl} r_m}{l^3}-\frac{{dQ} Q}{r_m}+\frac{3{dr_m} J^2}{2 r_m^4}\right)\epsilon\nonumber\\
&+&\left(\frac{2 {dr_m}}{l^2}+\frac{{dr_m} Q^2}{2 r_m^2}\right)\epsilon
+O\left(\epsilon^2\right).\label{eedm}
\ena
Substituting Eq.(\ref{eedm}) into Eq.(\ref{eedf2}), we can get lastly
\bea\label{eermdf}
dF_m
&=& \frac{{dr_m} \left(J^2-{4 r_m^4}/{l^2}+Q^2 r_m^2\right)}{2 r_m^3}\nonumber\\
&+& \left(\frac{{dJ} J}{r_m^3}+\frac{4{dl} r_m}{l^3}+\frac{{dQ} Q}{r_m}\right)\epsilon\nonumber\\
&+&{dr_m} \left(-\frac{3 J^2+Q^2 r_m^2}{2 r_m^4}-\frac{2}{l^2}\right)\epsilon\nonumber\\
&+&O\left(\epsilon^2\right).
\ena
Substituting Eq.(\ref{jn}) into  Eq.(\ref{eermdf}) further, we find
\bea\label{dfrm1}
dF_m
&=&\left(\frac{{dJ} \sqrt{4 r_m^4-l^2 Q^2 r_m^2}}{l r_m^3}+\frac{4 {dl} r_m}{l^3}+\frac{{dQ} Q}{r_m}\right)\epsilon\nonumber\\
&-&\left(\frac{8 {dr_m}}{l^2}-\frac{{dr_m} Q^2}{r_m^2}\right)\epsilon
\nonumber\\
&+&O\left(\epsilon^2\right).
\ena
In addition,  with Eq.(\ref{pf}), we can get
\be \label{eee}
dF'_{m}=\frac{\partial F'_{m}}{\partial J}dJ+\frac{\partial F'_{m}}{\partial Q}dQ+\frac{\partial F'_{m}}{\partial l}dl+\frac{\partial F'_{m}}{\partial r_{m}}dr_{m}=0.
\ee
Solving this equation, we obtain
\be \label{dJ}
dJ=-\frac{r_m \left(4 {dl} r_m^3+{dQ} l^3 Q r_m+{dr_m} l^3 Q^2-8 {dr_m} l r_m^2\right)}{l^2 \sqrt{4 r_m^4-l^2 Q^2 r_m^2}}.
\ee
Substituting Eq.(\ref{dJ})  into Eq.(\ref{dfrm1}), we  find
\bea\label{dfrmf}
dF_m=
O(\epsilon)^2.
\ena
Similar to the case without pressure, we can not ignore $O(\epsilon)^2$ for $\delta$ is also a small quantity. To  determine which is smaller, we should find out the coefficient of  $\epsilon^2$.

\begin{figure}[H]
\centering
\includegraphics[scale=0.55]{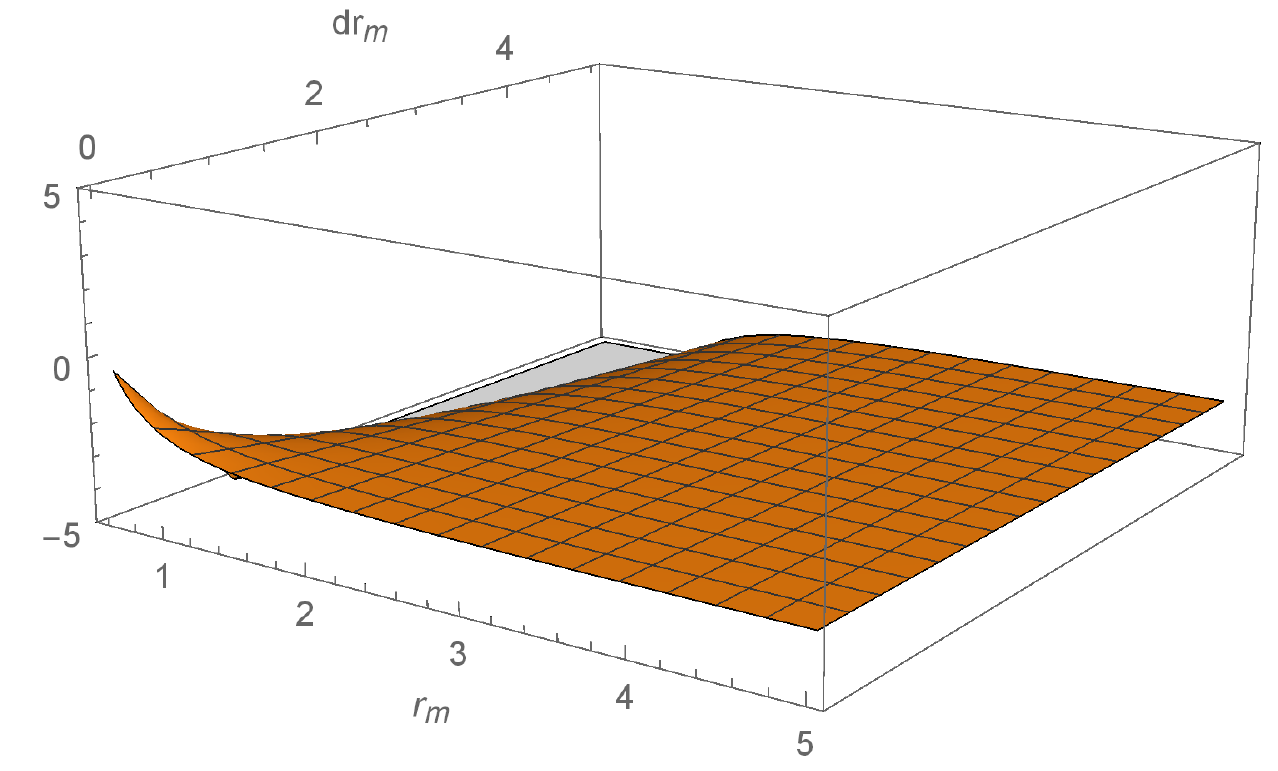}
 \caption{\small The values  of $\bar F$ for different  $r_m$ and $dr_m$  for the case $  l = Q=1, dQ=0.5, dl=0.1$.   } \label{fig4}
\end{figure}

\begin{figure}[H]
\centering
\includegraphics[scale=0.55]{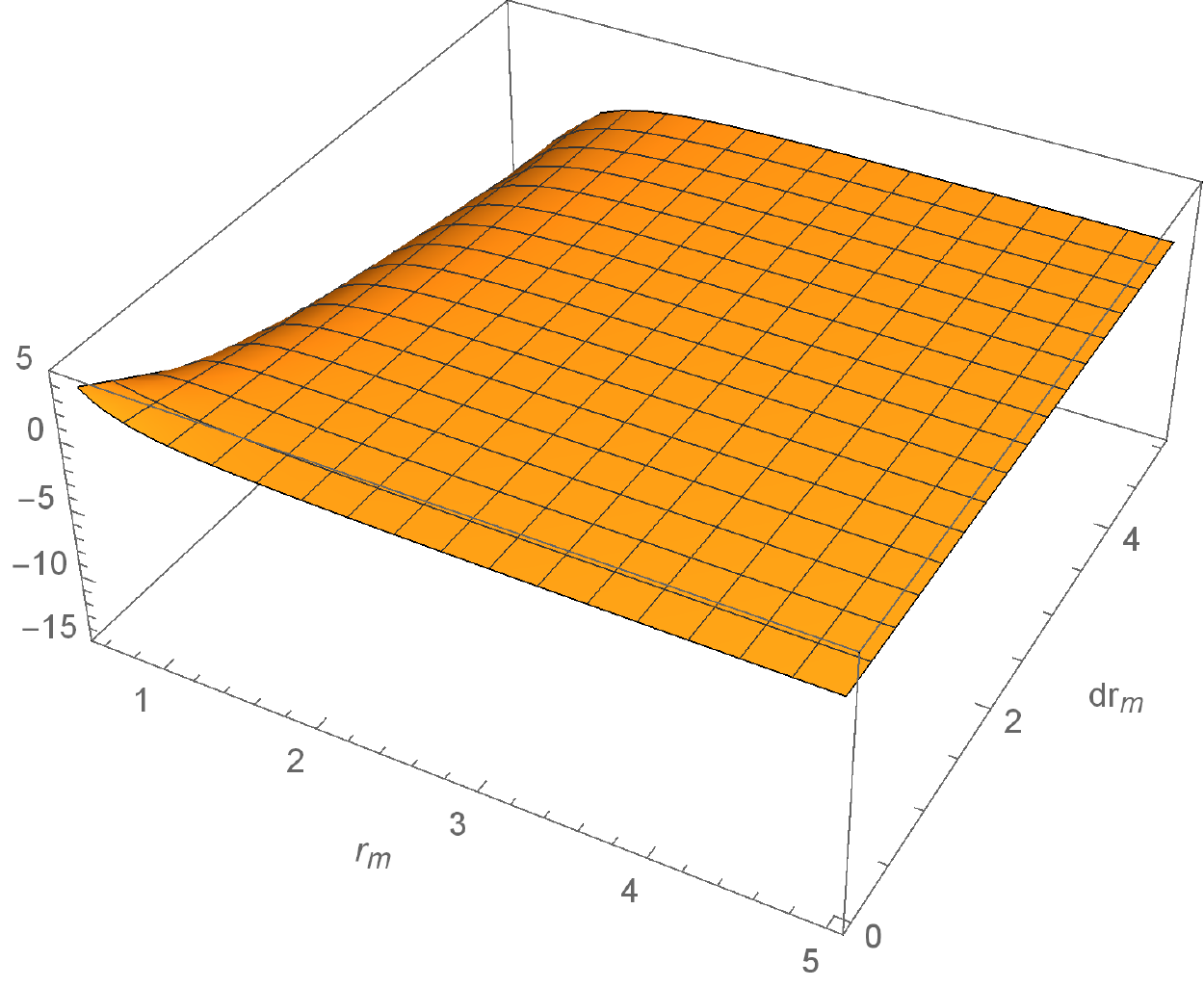}
 \caption{\small The values  of $\bar F$ for different  $r_m$ and $dr_m$  for the case $  l = Q=1, dQ=0.5, dl=0.6$.   } \label{fig5}
\end{figure}

As we expand $dM$ to the second order, we find
\be \label{dffinal}
dF_m= \left(\frac{8  {dl}}{l^3}+\frac{Q \left( {dQ} r_m- {dr_m} Q\right)}{r_m^3}\right)\epsilon^2 +O\left(\epsilon^3\right).
\ee
Based on   Eq.(\ref{deltaf}) and   Eq.(\ref{dffinal}), we find
\bea
\bar F&\equiv&F(r_m+dr_m)/\epsilon^2 \nonumber \\
&=&\frac{8  {dl}}{l^3}+\frac{Q \left( {dQ} r_m- {dr_m} Q\right)}{r_m^3}-\frac{4}{l^2}+\frac{Q^2}{2 r_m^2}.
\ena
It is obvious that the value of $\bar F$ depends on the values of $dl, l,dQ, Q, r_m, dr_m $
But we  find $dl$ affect  $\bar F$  drastically. For the case  $dl=0.1$, the value of $\bar F$ is negative, which is shown in Fig. \ref{fig4}. And for the case $dl=0.6$, the value of $\bar F$ is positive, which is shown in Fig. \ref{fig5}. Basically, $\bar F$  will be larger as we increase the value of  $dl$. The critical value of $dl$ which divides $\bar F$ into negative region and positive region is
\be
dl_c=\frac{l \left(-2  {dQ} l^2 Q r_m+2 {dr_m} l^2 Q^2-l^2 Q^2 r_m+8 r_m^3\right)}{16 r_m^3}.
\ee

In a word, from   Fig. \ref{fig5}, we know that the minimum  value of function $\bar F$ may be positive at the final state  for the near-extremal BTZ black hole
 in the extended phase space with pressure.
 So the
weak cosmic censorship conjecture may be violated for the near-extremal BTZ black hole in the extended phase space  with pressure, which is different from the previous observation.

 \section{ Conclusions}\label{6}
 As a charged spinning fermion is absorbed by a charged rotating BTZ black hole, we investigated the dynamics of the fermion and the variations of the thermodynamical quantities  of the black hole. It was found that the momentum of the particle was related to the chemical potential and angular velocity of the black hole. As the energy conservation, charge conservation and angular momentum conservation are considered, this relation was found to be nothing but the first law of thermodynamics for cases with and without pressure respectively.

We also investigated the second law of thermodynamics by discussing the variation of entropy. For the case without pressure, we found the variation of entropy was positive always for the non-extremal BTZ black hole, implying that the second law is valid in this case. While in the extended phase space  with pressure, because of the existence of $dl, dQ$, we found that the variation of entropy may be negative, depending on the values of  $l, Q$ and their variations. The second law thus may be violated in the extended phase space with pressure.

The absorbed fermion would change the location of the event horizon, thus we also discussed the weak cosmic censorship conjecture by discussing the variation of the minimum value of the function  $F(r)$. In the normal phase space, we found that the  minimum value of the function $F(r)$ did not change for the extremal BTZ black hole while became smaller   for the near-extremal BTZ black hole.  The final states of the near-extremal black holes thus have  solutions always. Our results showed that the weak cosmic censorship conjecture was valid without pressure since  the final states for both the minimum values of the  extremal and near-extremal black holes have solutions. In the extended phase space with pressure, we found that the function $F(r)$ did not change too for the extremal BTZ  black hole. However for the near-extremal BTZ  black hole, the function $F(r)$   became larger for some parameters.  The final states of the near-extremal black hole in this case does not have solutions. Thus we concluded that the weak cosmic censorship conjecture was violable as the pressure is considered, which is different from the previous result where  the weak cosmic censorship conjecture was found to be valid \cite{Gwak:2017kkt}. The reason arises from that the second order term of $dF_m$ was neglected previously. However, the initial states of the near-extremal black holes are also small quantities, the second order terms thus can not be neglected directly. It seems that our conclusions are more reasonable.

\section*{Acknowledgements}{ This work is supported  by the National
Natural Science Foundation of China (Grant No. 11875095), and Basic Research Project of Science and Technology Committee of Chongqing (Grant No. cstc2018jcyjA2480).}

\end{document}